\documentclass[cameraready]{Interspeech}

\title{Speech Encoder Fusion for LLM-based Automatic Speech Recognition}

\author{Jakob}{Poncelet}
\author{Hugo}{Van hamme}
\address{
    KU Leuven \\ Department Electrical Engineering ESAT-PSI, Leuven, Belgium
}
\email{}


\usepackage{amsmath,graphicx,hyperref,booktabs,multirow}

\begin{document}
\maketitle

\begin{abstract}
Speech-aware large language models (LLMs) can incorporate speech through pre-trained acoustic encoders that project speech features into the LLM embedding space. While the choice of the speech encoder critically influences performance, different encoders often exhibit complementary strengths, motivating their combination. In this work, we investigate whether fusing multiple pre-trained speech encoders can enhance speech-aware LLMs for automatic speech recognition (ASR). We explore several fusion strategies beyond simple feature concatenation, including learned combinations and Transformer-based fusion architectures, and evaluate them across mono- and multilingual ASR settings as well as diarized speech recognition. Our results indicate that carefully fusing multiple parallel speech encoders improves downstream performance in all scenarios with limited computational overhead.
\end{abstract}

\section{Introduction}
\label{sec:intro}

Recent advancements have led to the development of multimodal LLMs that can process other modalities besides text, such as images or audio \cite{blip2,llava,qwen_audio,salmonn}. These systems typically leverage a modality-specific encoder which converts the input image or audio to feature vectors, which are then aligned with the embedding space of the LLM via an adapter or connector module. Augmenting large language models (LLMs) to directly process speech signals has attracted significant attention, driving the rapid development of speech-aware LLMs \cite{fathullah-etal-2024-audiochatllama, seide24_interspeech, deng-etal-2025-wav2prompt, chen24_slt}. In such speech-augmented LLMs, it is an acoustic encoder that extracts the relevant speech features which are projected to the embedding space of the LLM \cite{ma25_aaai,yu2023connecting,verdini2024connect,prompting2024}. 

Typically, this speech encoder is pre-trained on large-scale speech data in a supervised or self-supervised manner \cite{whisper,owsm,wav2vec2,wavlm}, such that only the adapter layer and possibly the LLM have to be finetuned. As a result, the choice of the encoder has a big impact on the performance of the resulting system \cite{verdini2024connect}.

This work explores whether multiple speech encoders can be combined to create more robust speech-aware LLMs. We focus on a single task, specifically Automatic Speech Recognition (ASR), in a very common setting: there is a large open-source multilingual speech model available (e.g. Whisper \cite{whisper}) as well as a dedicated monolingual speech model that might be trained or finetuned on a specific domain. Generally, while one model might attain better WERs than the other one on a specific test set or for a specific language, the mistakes they make are often not the same and the predictions are complementary \cite{prakash25_interspeech}. Hence, we assume that a combination of the speech models can be beneficial, leveraging the complementary strengths of multiple speech encoders. Given their non-autoregressive, parallel nature \cite{vaswani}, encoders are substantially faster than autoregressive decoders, meaning parallelizing multiple encoders incurs minimal overhead.

We explore several methods to fuse the features of the pre-trained speech encoders during finetuning of the LLM for the task of ASR. More specifically, we analyze whether additional strategies besides simple feature concatenation are beneficial for mono- and multilingual speech recognition, such as learned combinations or fusion Transformer architectures.

We evaluate the possibilities for encoder fusion on four tasks. First, we focus on (Belgian) Dutch speech recognition, combining Whisper's encoder \cite{whisper} with the encoder of a Dutch-only ASR model \cite{jakob}. The Dutch-specific ASR model is an order of magnitude smaller than the Whisper model and pre-trained on much less data. Additionally, we experiment with English speech recognition, combining Whisper \cite{whisper} with a finetuned Wav2Vec2 model \cite{wav2vec2}. Second, we evaluate whether fusion of the encoders of two models is a viable approach for multilingual use cases. To this end, we explore training on two languages simultaneously and analyze whether fusing a matched monolingual model with a large multilingual model attains the desired benefits of getting improvements in the target language but still retains the capabilities of the multilingual model. For these experiments, we combine English and Dutch data and models. Third, we evaluate fusion of a speech encoder with a (pre-trained) speaker encoder for diarized speech recognition. Finally, we analyze the effect of having the ASR predictions in addition to the speech encoder features in the LLM.

\section{Related Work}
Most research on building speech-aware LLMs with pre-trained speech and text models focuses on single encoder systems \cite{ma25_aaai,yu2023connecting,verdini2024connect}. Previous works have leveraged a combination of multiple encoders, combining Whisper with WavLM \cite{hu-etal-2024-wavllm}, a speaker encoder \cite{lin2025diar}, or several weak encoders \cite{mowe_audio}, but they typically just concatenate or sum the output features. These works focus on multi-task speech processing in a single language (English), and do not target other languages. Alternative work has proposed a (linear) mixture of expert speech encoders conditioned on a representation of the task prompt \cite{shan2025enhancingspeechlargelanguage}.

Additionally, recent work \cite{prakash25_interspeech} has shown that combining predictions from multiple ASR systems with an LLM improves speech recognition. However, generating predictions with attention-based decoders is computationally heavy.

Our work differentiates from previous work by analyzing detailed fusion architectures (beyond concatenation/summing), focusing on a lower resourced language and speech recognition, and experimenting with multilingual fusion.

\section{Method}
We explore several methods to extend single-encoder speech LLMs to multi-encoder speech LLMs through encoder fusion.

\subsection{Speech LLM}
A speech encoder generates a sequence of feature vectors, which are downsampled and projected to the embedding dimension of the LLM by the projector. Then, the whole cascaded system of speech encoder, projector and LLM is finetuned on the task of speech recognition, where the LLM prompt consists of the projected speech features and a task description like \textit{"Transcribe the speech to text"}. 

In most works, the speech encoder is frozen and only the projector is trained in combination with low-rank updates of the LLM. This paper follows \cite{ma25_aaai} where downsampling of the speech features is done through stacking of $k=3$ consecutive feature vectors and the projection layer is a 2-layer MLP.

\subsection{Encoder Fusion}
\begin{figure}
    \centering
    \includegraphics[width=0.8\linewidth]{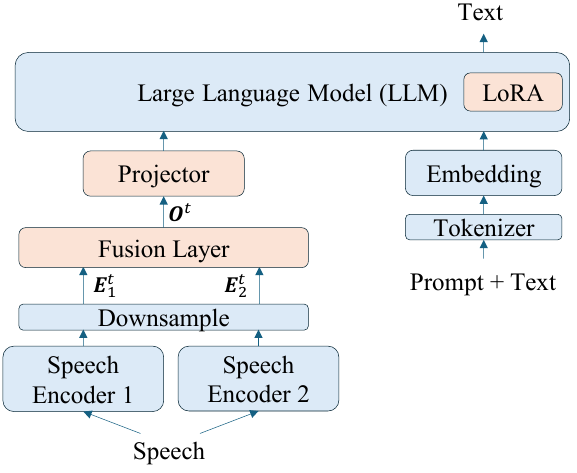}
    \caption{Speech-LLM with encoder fusion of multiple pre-trained speech encoders.}
    \label{fig:model}
\end{figure}

Multiple speech encoder outputs are combined with a fusion layer before being fed to the LLM projector, as shown in Figure~\ref{fig:model}. The output features of the speech encoders are denoted as $\textbf{E}_1^t$ and $\textbf{E}_2^t$, with $t=1 \dots T$. The sequence length $T$ of both encoder outputs is made equal by choosing the correct stacking (i.e. downsampling) factor $k$ for both. The resulting fused encoder outputs that are passed to the projection layer are denoted as $\textbf{O}^t$. 

Note that all proposed methods can be extended to more than two encoders. Crucially, no methods increase the speech vectors' sequence length, such that the context window of the LLM remains reasonably sized. 

\textbf{Feature concatenation:} The most straightforward way to combine two sequences of the same length is through concatenation of the features in the feature dimension at every time step, as proposed in \cite{hu-etal-2024-wavllm}.
\begin{equation}
    \mathbf{O}^t = \left[ \mathbf{E}_1^t; \mathbf{E}_2^t \right]
    \label{eq:concat}
\end{equation}

While the concatenation method is simple, it results in a rather static projection, which might not exploit encoder-specific strengths or be very adaptive to the input speech's language or domain.

\textbf{Sigmoid gate: } The encoder features are first linearly projected to the same inner dimension through transformations $\textbf{W}_1$ and $\textbf{W}_2$. Then, the output is constructed as a linear combination $\textbf{W}_a$ of both projected streams, using frame-specific gating weights $\alpha^t$. These gating weights are computed with a linear projection from the concatenation of the projected streams, followed by a sigmoid (or softmax for more than 2 encoders). As such, we attain a more principled weighting of the encoder features.
\begin{gather}
  \alpha^t
    = \sigma\mkern-3mu\left(
         \mathbf{W}_a \left[
           \mathbf{W}_1 \mathbf{E}_1^t ;
           \mathbf{W}_2 \mathbf{E}_2^t
         \right]
       \right) \\
  \mathbf{O}^t
    = \alpha^t \mathbf{W}_1 \mathbf{E}_1^t
       + \bigl(1 - \alpha^t\bigr)\mathbf{W}_2 \mathbf{E}_2^t
\end{gather}

\textbf{Multi-head gate: } Similar to the sigmoid gate, the encoder features are first projected to a common inner dimension. To fuse them, we apply Multi-Head Attention (MHA) independently at each time step, attending across the encoder streams rather than across time. For a given frame (time step), we compute a single query (q) vector by passing the concatenated encoder features through a linear layer. The key (k) and value (v) sequences consist of the individual projected encoder features at that exact same frame. This means the `sequence length' that the attention mechanism sees is simply the number of encoders, not the length of the audio.

Because the attention softmax is calculated over the encoders, the MHA dynamically assigns weights that sum to 1 across the different streams. This allows multiple attention heads to independently determine how much each encoder should contribute to the final fused representation at that specific moment. The motivation is that different heads can learn to prioritize specific encoder streams based on underlying data- or language-specific characteristics.
\begin{gather}
  \mathbf{Q}^t
    = \mathbf{W}_a \left[
         \mathbf{W}_1 \mathbf{E}_1^t ;
         \mathbf{W}_2 \mathbf{E}_2^t
       \right] \\
  \mathbf{O}^t
    = \mathrm{MHA}\mkern-3mu\left(
         q = \mathbf{Q}^t,\;
         k = v = \left[
           \mathbf{W}_1 \mathbf{E}_1^t ,
           \mathbf{W}_2 \mathbf{E}_2^t
         \right]
       \right)
\end{gather}

\textbf{Positional Transformer: } The encoder streams are concatenated in the feature dimension at every position and linearly projected to feature vectors that are passed through a Transformer encoder. Attending over the entire sequence could provide information on the language or domain to correctly weigh the importance of the encoders. The linear projection is useful to perform dimensionality reduction.
\begin{gather}
    \mathbf{Q}^t
    = \mathbf{W}_a \left[
         \mathbf{E}_1^t ;
         \mathbf{E}_2^t
       \right] \\
    \mathbf{O}
    = \mathrm{MHA}\mkern-3mu\left(
         q = k = v = \mathbf{Q} 
        \right)
\end{gather}

\textbf{Temporal Transformer: } Instead of concatenating the features in the feature dimension, we combine the encoder streams in the temporal dimension. To this end, the encoder streams are first linearly mapped to an inner dimension, and then the projected feature vectors are interleaved to create a sequence of length $2*T$.\footnote{In our experiments, interleaving, i.e. $\left[\textbf{E}_1^1, \textbf{E}_2^1, \textbf{E}_1^2, \textbf{E}_2^2, ...\right]$, worked much better than concatenating them, i.e. $\left[\textbf{E}_1^1, \textbf{E}_1^2, ..., \textbf{E}_2^1, \textbf{E}_2^2, ...\right]$.}
A Transformer encoder processes the interleaved features and the outputs are downsampled back to the original sequence length with a mean pooling layer. The window and stride are equal to the number of encoders. This setup has the advantage that the features have not been concatenated in the feature dimension and linearly projected before the Transformer layer(s).
\begin{align}
    \mathbf{Q}
    &= \left[ \textbf{W}_1 \mathbf{E}_1^1, \textbf{W}_2 \mathbf{E}_2^1, ..., \textbf{W}_1 \mathbf{E}_1^T, \textbf{W}_2 \mathbf{E}_2^T 
    \right] \\
    \mathbf{O}
    &= \mathrm{Pool}\mkern-3mu\left( \mathrm{MHA}\mkern-3mu\left(
         q = k = v = \mathbf{Q} 
        \right)
        \right)
\end{align}

\section{Setup} \label{sec:setup}

\hspace{10pt} \textbf{Speech encoder(s):} We use the \textit{Whisper-large-v3} encoder \cite{whisper} for all experiments. For Dutch experiments, we incorporate the encoder from the \textit{NeLF} ASR model\footnote{HF \url{nelfproject/NeLF_S2T_Pytorch}} \cite{jakob}, which is a Conformer speech encoder pre-trained on 14k hours of weakly supervised Belgian Dutch speech in an encoder-decoder model with CTC regularization \cite{hybridctcatt} and two distinct decoders, where one is trained with subtitle data and the other is trained with verbatim data \cite{jakobslt}. For English, we use a CTC-finetuned Wav2vec2 model\footnote{HF \url{facebook/wav2vec2-large-robust}} pre-trained on a large variety of English data and finetuned on Librispeech \cite{librispeech}. For diarization, we use the pre-trained ECAPA2 model\footnote{HF \url{Jenthe/ECAPA2}} as speaker encoder \cite{ecapa2}, where the features are extracted before the pooling layers. 

\textbf{Downsampling:} All speech encoder features are downsampled to 16.7 Hz, i.e. one vector every 60 ms, through stacking. For \textit{Whisper}, this means stacking 3 consecutive features.

\textbf{Fusion layers:} We select the hidden dimensions of the linear projections and attention layers such that all methods have a similar number of parameters. All MHA layers use 4 heads.

\textbf{Projector:} The projector is a simple 2-layer MLP with inner dimension of 2048, ReLU, and outer dimension of 4096.

\textbf{LLM:} For Dutch experiments, we use \textit{Tweety-7B} \cite{remy24_colm}, which is a Dutch adaptation of \textit{Mistral-7B}. For English and multilingual experiments, we use \textit{Llama-3.1-8B}. Both LLMs are base models. The LLMs are quantized to 4-bit and finetuned using QLoRA for efficient training. We use rank 4, alpha 16 and dropout 0.05 for QLoRA applied to all linear mappings (attention and feedforward) in the LLM. Speech is presented to the LLM before the task in the prompt, following \cite{fan25_jstsp}.

\textbf{Optimization:} Models are trained until convergence for a maximum of 5 epochs with a linear decaying learning rate with peak value of 5e-5, 10\% warm-up, and an effective batch size of 128 utterances using the 8bit Adam optimizer. There are ~30M trainable parameters in total (fusion layer, projector and QLoRA), not taking into account quantization.

\textbf{Data:} For Dutch experiments, we train on 240 hours of Belgian Dutch ASR data from the Spoken Dutch Corpus \cite{oostdijk-2000-spoken}. We evaluate on a \textit{clean} test set (8 hours) derived from held-out speakers of the same corpus as well as a \textit{other} test set (6 hours) with manual transcripts of broadcast media \cite{jakobslt}. For English experiments, we train on 960h of Librispeech \cite{librispeech} and evaluate on \textit{test-clean} and \textit{test-other}. For multilingual experiments, we combine Librispeech's \textit{train-clean-360h} with the Dutch data. 

\textbf{Evaluation:} We report normalized Word Error Rates (WER). Based on test set sizes and baseline error rates, we estimate a global maximum margin of error (at a 95\% confidence level) of $\leq \pm 0.1\%$ for standard evaluations and $\leq \pm 0.3\%$ for diarized evaluations using the standard proportion formula.

\section{Results}

\subsection{Monolingual speech recognition}
For Dutch experiments, we combine the encoders of the multilingual Whisper and the monolingual NeLF model using the proposed fusion mechanisms. For English, we combine Whisper and the monolingual (finetuned) Wav2vec2 model. Results on Dutch (NL) are in Table~\ref{tab:vl}, results on English are in Table~\ref{tab:eng}.

\begin{table}[t]
    \centering
    \footnotesize
    \caption{Monolingual ASR -- WER (\%) for Dutch experiments: training on Dutch CGN data and evaluating on Dutch (NL). The LLM is Tweety-7B.}
    \vspace{-7pt}
    \begin{tabular}{c|c|c|c c}
        \toprule
        \multicolumn{3}{c|}{\textbf{Model}} & \multicolumn{2}{c}{\textbf{WER on NL}} \\
        \textbf{Encoder(s)} & \textbf{Fusion mode} & \textbf{Decoder} & \textbf{\textit{clean}} & \textbf{\textit{other}} \\
        \midrule
        Whisper & / & \multirow{2}{*}{LLM} & 8.3 & 11.5 \\
        NeLF & / & & 7.5 & 9.0 \\
        \hline
        \multirow{5}{*}{\shortstack{Whisper\\+\\NeLF}} & Concat & \multirow{5}{*}{LLM} & 7.2 & 8.9 \\
         & Sigmoid gate &  & 7.1 & \underline{8.4} \\
         & Multi-head gate & & \underline{7.0} & 8.7 \\
         & Positional Transf. &  & 7.1 & 8.7 \\
         & Temporal Transf. & & \textbf{6.8} & \textbf{8.3} \\
        \bottomrule
    \end{tabular}
    \label{tab:vl}
\end{table}

\begin{table}[t]
    \centering
    \footnotesize
    \caption{Monolingual ASR -- WER (\%) for English experiments: training and evaluating on English LibriSpeech (EN). The LLM is LLama-3.1-8B.}
    \vspace{-7pt}
    \begin{tabular}{c|c|c|c c}
        \toprule
        \multicolumn{3}{c|}{\textbf{Model}} & \multicolumn{2}{c}{\textbf{WER on EN}} \\
        \textbf{Encoder(s)} & \textbf{Fusion mode} & \textbf{Decoder} & \textbf{\textit{clean}} & \textbf{\textit{other}} \\
        \midrule
        Whisper & / & \multirow{2}{*}{LLM} & 3.2 & 6.4 \\
        Wav2vec2-FT & / & & 3.5 & 6.0 \\
        \hline
        \multirow{5}{*}{\shortstack{Whisper\\+\\Wav2vec2-FT}} & Concat & \multirow{5}{*}{LLM} & 3.3 & 6.2 \\
         & Sigmoid gate & & \textbf{2.8} & \textbf{5.5} \\
         & Multi-head gate & & \underline{3.0} & 6.0 \\
         & Positional Transf. & & 3.5 & 6.2 \\
         & Temporal Transf. & & 3.1 & \underline{5.9} \\
        \bottomrule
    \end{tabular}
    \label{tab:eng}
    \vspace{-10pt}
\end{table}

For Dutch, we report nice improvements as a result of complementary strengths of Whisper and NeLF encoders. Note that Whisper's WER is worse than NeLF, but the combination is still helpful, likely e.g. for better prediction of rare words, words related to foreign languages, and difficult acoustic settings. While all methods outperform the concatenation baseline, the temporal transformer seems to capture complementary aspects best. For English, we noticed that obtaining well-converged systems was more difficult (probably related to Librispeech and LLM pre-training) in all settings. Simple sigmoid gating works best to combine the features of the two highly optimized models, whereas more difficult setups are probably unnecessary.

\subsection{Multilingual speech recognition}
We train a speech-LLM on the two languages jointly, i.e. Dutch and English. In this setup, the prompt is adapted to \textit{"Transcribe the speech to text in the same language."} Results are in Table~\ref{tab:ml}.

\begin{table}[t]
    \centering
    \footnotesize
    \caption{Multilingual ASR -- WER (\%) for joint experiments: training on combination of Dutch CGN data and English LibriSpeech-360h. Evaluating on Dutch test-clean (NL) and LibriSpeech test-clean (EN). The LLM is Llama-3.1-8B.}
    \vspace{-7pt}
    \begin{tabular}{c|c|c|c c}
        \toprule
        \multicolumn{3}{c|}{\textbf{Model}} & \multicolumn{2}{c}{\textbf{WER (clean)}} \\
        \textbf{Encoder(s)} & \textbf{Fusion mode} & \textbf{Decoder} & \textbf{\textit{NL}} & \textbf{\textit{EN}} \\
        \midrule
        Whisper & / & \multirow{2}{*}{LLM} & 8.4 & 2.9 \\
        NeLF & / & & 7.4 & 10.9 \\
        \hline
        \multirow{5}{*}{\shortstack{Whisper\\+\\NeLF}} & Concat & \multirow{5}{*}{LLM} & 7.1 & 3.9 \\
         & Sigmoid gate &  & \underline{6.6} & \underline{2.7} \\
         & Multi-head gate & & \textbf{6.5} & \textbf{2.5} \\
         & Positional Transf. &  & 6.8 & 3.0 \\
         & Temporal Transf. & & 6.7 & 3.1 \\
        \bottomrule
    \end{tabular}
    \label{tab:ml}
\end{table}

Fusion layers in multilingual settings offer strong benefits over concatenation as they facilitate language-dependent optimization more easily. In this setup, multi-head gating performs best, although the Transformer-based methods also improve recognition. We remark that multilingual training greatly improved convergence with the Whisper encoder and leads to more stable and better results, even for English.

\subsection{Diarized speech recognition}
We train the speech-LLM to return a diarized transcription, e.g. \textit{``0: Hello. 1: How are you? 0: I'm fine."}, on Dutch data. To this end, we combine the NeLF ASR encoder with the pre-trained ECAPA speaker encoder. As the speaker encoder is generally not trained to contain fine-grained temporal information, we do averaging instead of stacking when downsampling, which leads to more stable results and outputs.
As an additional baseline, we include the gated cross attention mechanism for fusion of a speech and a speaker encoder proposed in \cite{lin2025diar}. In this method, the speech features (queries) cross-attend to the speaker features (keys/values), and then the attended features are linearly gated and combined with the original speech features.

\noindent We train on the same Dutch CGN data, for which ~40\% of the examples are multi-speaker (with a max of 4 speakers). We evaluate on a multi-speaker-only subset of the test set and report several diarized ASR metrics. Note that the NeLF model was already pre-trained to output speaker-change symbols \textit{\textless spk\textgreater} with serialized output training \cite{sot}, so the benefit of the ECAPA model should be in fortifying these speaker detections as well as performing diarization of the same speakers within an utterance.
We report the Speaker-Attributed WER (SA-WER) \cite{fiscus2008}, which includes speaker labels in WER, counting a word as correctly recognized only if both the transcription and the assigned speaker identity match the reference.
To isolate diarization errors, we report the flat WER with pure transcription accuracy, and the Speaker Confusion (Spk-Conf) which measures the frequency of words that were transcribed correctly but attributed to the wrong speaker. Results are in Table~\ref{tab:diar}.

\begin{table}[t]
    \centering
    \footnotesize
    \caption{Diarized ASR -- Dutch experiments: training and evaluating on diarized Dutch CGN data. The test set is filtered to contain only multi-speaker utterances. The LLM is Tweety-7B.}
    \vspace{-7pt}
    \resizebox{\columnwidth}{!}{ %
    \begin{tabular}{c|c|c c c}
        \toprule
        \multicolumn{2}{c|}{\textbf{Model}} & \multicolumn{3}{c}{\textbf{Metrics (\%)}} \\
        \textbf{Encoder(s)} & \textbf{Fusion mode} & \textbf{SA-WER} & \textbf{WER} & \textbf{Spk-Conf} \\
        \midrule
        NeLF & / & 24.7 & 16.8 & 7.9 \\
        \hline
        \multirow{6}{*}{\shortstack{NeLF\\+\\ECAPA}} & 
          Cross-attention \cite{lin2025diar} & 22.6 & 16.3 & 6.3 \\
         & Concat & 21.4 & \underline{16.2} & 5.2 \\
         & Sigmoid gate & 23.4 & 16.3 & 7.1 \\
         & Multi-head gate & 26.8 & 21.0 & 5.8 \\
         & Positional Transf. & \underline{19.7} & 16.4 & \textbf{3.3} \\
         & Temporal Transf. & \textbf{18.1} & \textbf{14.5} & \underline{3.6} \\
        \bottomrule
    \end{tabular}
    }
    \label{tab:diar}
    \vspace{-10pt}
\end{table}

For diarized speech recognition, the transformer-based fusion methods excel at capturing long-range speaker feature similarity. Despite robust speaker separation of the source NeLF encoder, the additional speaker encoder reduces speaker confusion to a minimum without hindering recognition accuracy.

\subsection{Including decoder predictions in LLM prompt}
As common benchmarks have shown \cite{openasr}, it is often difficult to outperform a dedicated ASR model with a speech-LLM. Moreover, it is not always favorable to train the speech-LLM on as much data as the ASR models, due to resource constraints. Therefore, it can be beneficial to feed an initial prediction to the LLM in addition to the speech features. For speech model fusion, we first train the speech-LLM on ASR as before with only speech features. Then, in a second stage we finetune the model to transcribe the speech features given both the speech and some initial hypotheses. The hypotheses are computed with the decoders of the speech models used in the fusion. While this increases processing time, it can be worthwhile in offline settings. As an additional baseline, we include the result of finetuning and evaluating the LLM with text data only (i.e. the decoder predictions). Results for Dutch and English are in Table~\ref{tab:dec-vl} and~\ref{tab:dec-en}. 

\begin{table}[t]
    \centering
    \footnotesize
    \caption{Monolingual ASR with pre-trained decoder predictions -- WER (\%) for Dutch experiments: training and evaluating on Dutch data (NL). The LLM is Tweety-7B. For NeLF, \textit{verbatim} and \textit{subtitle} refer to the two decoders in the model (see Sec.~\ref{sec:setup}).}
    \vspace{-7pt}
    \begin{tabular}{c|c|c|c c}
        \toprule
        \multicolumn{2}{c}{\textbf{Model}} & \textbf{Included} & \multicolumn{2}{c}{\textbf{WER on NL}} \\
        \textbf{Encoder(s)} & \textbf{Decoder} & \textbf{Predictions} & \textbf{\textit{clean}} & \textbf{\textit{other}} \\
        \midrule
        Whisper & Whisper & / & 11.3 & 13.1 \\
        NeLF & NeLF & / & 6.8 & 8.2 \\
        \hline
        Whisper & \multirow{2}{*}{LLM} & / & 8.3 & 11.5 \\
        NeLF & & / & 7.5 & 9.0 \\
        \hline
        \multirow{5}{*}{Fusion} & \multirow{5}{*}{LLM} & / & 6.8 & 8.3 \\
        & & NeLF verbatim & 6.4 & 7.8 \\
        & & NeLF subtitle & \underline{5.9} & \textbf{7.5} \\
        & & Whisper output & 6.0 & \underline{7.7} \\
        & & All & \textbf{5.6} & 7.8 \\
        \hline
        / (text-only) & LLM & All & 6.0 & 8.1 \\
        \bottomrule
    \end{tabular}
    \label{tab:dec-vl}
\end{table}

\begin{table}[t]
    \centering
    \footnotesize
    \caption{Monolingual ASR with pre-trained decoder predictions -- WER (\%) for English experiments: training and evaluating on Librispeech (EN). The LLM is Llama-3.1-8B.}
    \vspace{-7pt}
    \begin{tabular}{c|c|c|c c}
        \toprule
        \multicolumn{2}{c}{\textbf{Model}} & \textbf{Included} & \multicolumn{2}{c}{\textbf{WER on EN}} \\
        \textbf{Encoder(s)} & \textbf{Decoder} & \textbf{Predictions} & \textbf{\textit{clean}} & \textbf{\textit{other}} \\
        \midrule
        Whisper & Whisper & / & 2.0 & 3.7 \\
        Wav2vec2-FT & CTC & / & 2.6 & 5.3 \\
        \hline
        Whisper & \multirow{2}{*}{LLM} & / & 3.2 & 6.4 \\
        Wav2vec2-FT & & / & 3.5 & 6.0 \\
        \hline
        \multirow{4}{*}{Fusion} & \multirow{4}{*}{LLM} & / & \underline{2.8} & \underline{5.5} \\
        & & Whisper & 3.6 & 5.8 \\
        & & Wav2vec2-FT & 3.9 & 6.2 \\
        & & All & \textbf{2.1} & \textbf{3.8} \\
        \hline
        / (text-only) & LLM & All & 1.4 & 3.1 \\
        \bottomrule
    \end{tabular}
    \label{tab:dec-en}
    \vspace{-10pt}
\end{table}

For Dutch, speech-only LLM with fusion layers matches the best ASR model, but including decoder predictions maximizes the speech-LLM performance and strongly outperforms both baseline ASR models and the text-only (error-correcting) LLM. For English, it is more difficult to match the ASR models with speech-LLMs in our setup, although using the predictions can close the gap. The text-based LLM outperforms the speech-LLMs, which might be related to the text origin of Librispeech that is present in the LLM pre-training data and that no trained parameters are used towards interpreting the speech signal \cite{tseng2025evaluationllmsspeechflawed}.

\section{Discussion}
Encoder fusion for speech-LLM is a simple technique with limited overhead to incorporate strengths from multiple pre-trained speech encoders. Although speech-LLMs have difficulties attaining performance of dedicated ASR systems, augmenting LLMs with speech capabilities has much wider applications besides ASR. Note that all experiments in this paper involve short-form speech recognition. Additional benefits with LLM-ASR can be attained by leveraging history text, larger ranks and unquantized LLMs (as this paper uses rank 4 and 4-bit quantization due to resource constraints) to attain better absolute WERs. 

\section{Conclusion}
We have explored several methods to fuse the outputs of multiple pre-trained speech encoders to attain stronger speech-LLM systems in a variety of scenarios. We found that careful fusion outperforms standard feature concatenation in all cases.

\newpage
\section{Acknowledgments}
\ifcameraready
Research supported by the Research Foundation Flanders (FWO) under grant S004923N of the SBO programme and by the Flemish Government under the "Flanders AI Research Program". Part of the resources and services used in this work were provided by the VSC (Flemish Supercomputer Center), funded by the Research Foundation Flanders (FWO) and the Flemish Government.
\else
Anonymous.
\fi

\section{Use of Generative AI Disclosure}
The authors used Generative AI exclusively for text formatting, editing, and polishing to improve the clarity of this manuscript. No part of the manuscript's content or ideas was produced by generative AI. All authors take full responsibility and accountability for the original work and content of this paper.

\bibliographystyle{IEEEtran}
\bibliography{refs}

@inproceedings{blip2,
author = {Li, Junnan and Li, Dongxu and Savarese, Silvio and Hoi, Steven},
title = {{BLIP-2}: Bootstrapping language-image pre-training with frozen image encoders and large language models},
year = {2023},
booktitle = {Proc.  Conf. on Machine Learning (ICML)},
pages={19730--19742},
}

@inproceedings{llava,
 author = {Liu, Haotian and Li, Chunyuan and Wu, Qingyang and Lee, Yong Jae},
 booktitle = {Advances in Neural Information Processing Systems},
 pages = {34892--34916},
 title = {Visual Instruction Tuning},
 volume = {36},
 year = {2023}
}

@misc{qwen_audio,
      title={{Qwen-Audio}: Advancing Universal Audio Understanding via Unified Large-Scale Audio-Language Models}, 
      author={Yunfei Chu and Jin Xu and Xiaohuan Zhou and Qian Yang and Shiliang Zhang and Zhijie Yan and Chang Zhou and Jingren Zhou},
      year={2023},
      note={arXiv eprint 2311.07919}, 
}

@inproceedings{salmonn,
  title={{SALMONN}: Towards Generic Hearing Abilities for Large Language Models},
  author={Changli Tang and Wenyi Yu and Guangzhi Sun and Xianzhao Chen and Tian Tan and Wei Li and Lu Lu and Zejun MA and Chao Zhang},
  booktitle={Proc. Intl. Conf. on Learning Representations (ICLR)},
  year={2024},
}

@INPROCEEDINGS{yu2023connecting,
  author={Yu, Wenyi and Tang, Changli and Sun, Guangzhi and Chen, Xianzhao and Tan, Tian and Li, Wei and Lu, Lu and Ma, Zejun and Zhang, Chao},
  booktitle={Proc. IEEE Intl. Conf. on Acoustics, Speech and Signal Processing (ICASSP)}, 
  title={Connecting Speech Encoder and Large Language Model for {ASR}}, 
  year={2024},
  pages={12637-12641},
  doi={10.1109/ICASSP48485.2024.10445874},
}

@inproceedings{verdini2024connect,
  title     = {How to Connect Speech Foundation Models and Large Language Models? {What} Matters and What Does Not},
  author    = {Francesco Verdini and Pierfrancesco Melucci and Stefano Perna and Francesco Cariaggi and Marco Gaido and Sara Papi and Szymon Mazurek and Marek Kasztelnik and Luisa Bentivogli and Sebastien Bratières and Paolo Merialdo and Simone Scardapane},
  year      = {2025},
  booktitle = {Proc. Interspeech},
  pages     = {1813--1817},
  doi       = {10.21437/Interspeech.2025-2245},
}

@InProceedings{whisper,
  title = 	 {Robust Speech Recognition via Large-Scale Weak Supervision},
  author =       {Radford, Alec and Kim, Jong Wook and Xu, Tao and Brockman, Greg and Mcleavey, Christine and Sutskever, Ilya},
  booktitle = 	 {Proc. Intl. Conf. on Machine Learning (ICML)},
  pages = 	 {28492--28518},
  year = 	 {2023},
  volume = 	 {202},
}

@inproceedings{wav2vec2,
 author = {Baevski, Alexei and Zhou, Yuhao and Mohamed, Abdelrahman and Auli, Michael},
 booktitle = {Advances in Neural Information Processing Systems},
 pages = {12449--12460},
 title = {wav2vec 2.0: A Framework for Self-Supervised Learning of Speech Representations},
 volume = {33},
 year = {2020}
}

@inproceedings{hu-etal-2024-wavllm,
    title = "{W}av{LLM}: Towards Robust and Adaptive Speech Large Language Model",
    author = "Hu, Shujie  and
      Zhou, Long  and
      Liu, Shujie  and
      Chen, Sanyuan  and
      Meng, Lingwei  and
      Hao, Hongkun  and
      Pan, Jing  and
      Liu, Xunying  and
      Li, Jinyu  and
      Sivasankaran, Sunit  and
      Liu, Linquan  and
      Wei, Furu",
    booktitle = "Findings of the Association for Computational Linguistics: EMNLP",
    year = "2024",
    doi = "10.18653/v1/2024.findings-emnlp.263",
    pages = "4552--4572",
}

@INPROCEEDINGS{mowe_audio,
  author={Zhang, Wenyu and Sun, Shuo and Wang, Bin and Zou, Xunlong and Liu, Zhuohan and He, Yingxu and Lin, Geyu and Chen, Nancy F. and Ti Aw, Ai},
  booktitle={Proc. IEEE Intl. Conf. on Acoustics, Speech and Signal Processing (ICASSP)}, 
  title={{MoWE-Audio}: Multitask Audio{LLM}s with Mixture of Weak Encoders}, 
  year={2025},
  doi={10.1109/ICASSP49660.2025.10888128}}

@inproceedings{shan2025enhancingspeechlargelanguage,
    title = "Enhancing Speech Large Language Models with Prompt-Aware Mixture of Audio Encoders",
    author = "Shan, Weiqiao  and
      Li, Yuang  and
      Zhang, Yuhao  and
      Luo, Yingfeng  and
      Xu, Chen  and
      Zhao, Xiaofeng  and
      Meng, Long  and
      Lu, Yunfei  and
      Zhang, Min  and
      Yang, Hao  and
      Xiao, Tong  and
      Zhu, JingBo",
    booktitle = "Proc. Conf. on Empirical Methods in Natural Language Processing",
    doi = "10.18653/v1/2025.emnlp-main.974",
    year= "2025",
    pages = "19305--19320",
}

@inproceedings{ma25_aaai,
author = {Ma, Ziyang and Yang, Guanrou and Yang, Yifan and Gao, Zhifu and Wang, Jiaming and Du, Zhihao and Yu, Fan and Chen, Qian and Zheng, Siqi and Zhang, Shiliang and Chen, Xie},
title = {Speech recognition meets large language model: benchmarking, models, and exploration},
year = {2025},
doi = {10.1609/aaai.v39i23.34666},
booktitle = {Proc. AAAI Conf. on Artificial Intelligence},
articleno = {2768},
}

@misc{lin2025diar,
      title={Diarization-Aware Multi-Speaker Automatic Speech Recognition via Large Language Models}, 
      author={Yuke Lin and Ming Cheng and Ze Li and Beilong Tang and Ming Li},
      year={2025},
      eprint={2506.05796},
      archivePrefix={arXiv},
      primaryClass={eess.AS},
      note={arXiv eprint 2506.05796},
}

@InProceedings{fiscus2008,
author="Fiscus, Jonathan G.
and Ajot, Jerome
and Garofolo, John S.",
title="The Rich Transcription 2007 Meeting Recognition Evaluation",
booktitle="Multimodal Technologies for Perception of Humans",
year="2008",
pages="373--389",
}

@ARTICLE{wavlm,
  author={Chen, Sanyuan and Wang, Chengyi and Chen, Zhengyang and Wu, Yu and Liu, Shujie and Chen, Zhuo and Li, Jinyu and Kanda, Naoyuki and Yoshioka, Takuya and Xiao, Xiong and Wu, Jian and Zhou, Long and Ren, Shuo and Qian, Yanmin and Qian, Yao and Wu, Jian and Zeng, Michael and Yu, Xiangzhan and Wei, Furu},
  journal={IEEE Journal of Selected Topics in Signal Processing}, 
  title={{WavLM}: Large-Scale Self-Supervised Pre-Training for Full Stack Speech Processing}, 
  year={2022},
  volume={16},
  number={6},
  pages={1505-1518},
  doi={10.1109/JSTSP.2022.3188113}}

@inproceedings{owsm,
  title     = {{OWSM} v4: Improving Open Whisper-Style Speech Models via Data Scaling and Cleaning},
  author    = {Yifan Peng and Muhammad Shakeel and Yui Sudo and William Chen and Jinchuan Tian and Chyi-Jiunn Lin and Shinji Watanabe},
  year      = {2025},
  booktitle = {Proc. Interspeech},
  pages     = {2225--2229},
  doi       = {10.21437/Interspeech.2025-1062},
}

@INPROCEEDINGS{prompting2024,
  author={Fathullah, Yassir and Wu, Chunyang and Lakomkin, Egor and Jia, Junteng and Shangguan, Yuan and Li, Ke and Guo, Jinxi and Xiong, Wenhan and Mahadeokar, Jay and Kalinli, Ozlem and Fuegen, Christian and Seltzer, Mike},
  booktitle={Proc. IEEE Intl. Conf. on Acoustics, Speech and Signal Processing (ICASSP)}, 
  title={Prompting Large Language Models with Speech Recognition Abilities}, 
  year={2024},
  pages={13351-13355},
  doi={10.1109/ICASSP48485.2024.10447605}}

@inproceedings{prakash25_interspeech,
  title     = {Better Pseudo-labeling with Multi-{ASR} Fusion and Error Correction by Speech{LLM}},
  author    = {Jeena Prakash and Blessingh Kumar and Kadri Hacioglu and Bidisha Sharma and Sindhuja Gopalan and Malolan Chetlur and Shankar Venkatesan and Andreas Stolcke},
  year      = {2025},
  booktitle = {Proc. Interspeech},
  pages     = {579--583},
  doi       = {10.21437/Interspeech.2025-1707},
}

@inproceedings{vaswani,
 author = {Vaswani, Ashish and Shazeer, Noam and Parmar, Niki and Uszkoreit, Jakob and Jones, Llion and Gomez, Aidan N and Kaiser, \L ukasz and Polosukhin, Illia},
 booktitle = {Advances in Neural Information Processing Systems},
 title = {Attention is All you Need},
 volume = {30},
 year = {2017}
}

@article{jakob,
    author = {Poncelet, Jakob and Van hamme, Hugo},
    title = {Leveraging broadcast media subtitle transcripts for automatic speech recognition and subtitling},
    journal = {Journal on Audio, Speech and Music Processing},
    year = {2026},
    doi = {10.1186/s13636-026-00450-9},
    volume = {11},
}

@inproceedings{fathullah-etal-2024-audiochatllama,
    title = "{A}udio{C}hat{L}lama: Towards General-Purpose Speech Abilities for {LLM}s",
    author = "Fathullah, Yassir  and
      Wu, Chunyang  and
      Lakomkin, Egor  and
      Li, Ke  and
      Jia, Junteng  and
      Shangguan, Yuan  and
      Mahadeokar, Jay  and
      Kalinli, Ozlem  and
      Fuegen, Christian  and
      Seltzer, Mike",
    booktitle = "Proc. Conf. of the North American Chapter of the Association for Computational Linguistics (NAACL): Human Language Technologies (Volume 1: Long Papers)",
    year = "2024",
    doi = "10.18653/v1/2024.naacl-long.309",
    pages = "5522--5532",
}

@ARTICLE{fan25_jstsp,
  author={Fan, Ruchao and Ren, Bo and Hu, Yuxuan and Zhao, Rui and Liu, Shujie and Li, Jinyu},
  journal={IEEE Journal of Selected Topics in Signal Processing}, 
  title={{AlignFormer}: Modality Matching Can Achieve Better Zero-Shot Instruction-Following Speech-{LLM}}, 
  year={2025},
  volume={19},
  number={7},
  pages={1329-1337},
  doi={10.1109/JSTSP.2025.3588378}}

@inproceedings{seide24_interspeech,
  title     = {Speech {ReaLLM} – Real-time Speech Recognition with Multimodal Language Models by Teaching the Flow of Time},
  author    = {Frank Seide and Yangyang Shi and Morrie Doulaty and Yashesh Gaur and Junteng Jia and Chunyang Wu},
  year      = {2024},
  booktitle = {Proc. Interspeech},
  pages     = {1900--1904},
  doi       = {10.21437/Interspeech.2024-571},
}

@inproceedings{deng-etal-2025-wav2prompt,
    title = "{W}av2{P}rompt: End-to-End Speech Prompt Learning and Task-based Fine-tuning for Text-based {LLM}s",
    author = "Deng, Keqi  and
      Sun, Guangzhi  and
      Woodland, Phil",
    booktitle = "Proc. Conf. of the Nations of the Americas Chapter of the Association for Computational Linguistics (NAACL): Human Language Technologies (Volume 1: Long Papers)",
    year = "2025",
    doi = "10.18653/v1/2025.naacl-long.354",
    pages = "6940--6956",
}

@INPROCEEDINGS{chen24_slt,
  author={Chen, Zhehuai and Huang, He and Hrinchuk, Oleksii and Puvvada, Krishna C. and Koluguri, Nithin Rao and Żelasko, Piotr and Balam, Jagadeesh and Ginsburg, Boris},
  booktitle={Proc. IEEE Spoken Language Technology Workshop (SLT)}, 
  title={{BESTOW}: Efficient and Streamable Speech Language Model with The Best of Two Worlds in {GPT} and {T5}}, 
  year={2024},
  pages={147-154},
  doi={10.1109/SLT61566.2024.10832146}}

@ARTICLE{hybridctcatt,
  author={Watanabe, Shinji and Hori, Takaaki and Kim, Suyoun and Hershey, John R. and Hayashi, Tomoki},
  journal={IEEE Journal of Selected Topics in Signal Processing}, 
  title={Hybrid {CTC}/{A}ttention Architecture for End-to-End Speech Recognition}, 
  year={2017},
  volume={11},
  number={8},
  pages={1240-1253},
  doi={10.1109/JSTSP.2017.2763455}}

@INPROCEEDINGS{jakobslt,
  author={Poncelet, Jakob and Van Hamme, Hugo},
  booktitle={Proc. IEEE Spoken Language Technology Workshop (SLT)}, 
  title={Learning to Jointly Transcribe and Subtitle for End-To-End Spontaneous Speech Recognition}, 
  year={2023},
  pages={182-189},
  doi={10.1109/SLT54892.2023.10022420}}

@INPROCEEDINGS{librispeech,
  author={Panayotov, Vassil and Chen, Guoguo and Povey, Daniel and Khudanpur, Sanjeev},
  booktitle={Proc. IEEE Intl. Conf. on Acoustics, Speech and Signal Processing (ICASSP)}, 
  title={Librispeech: An {ASR} corpus based on public domain audio books}, 
  year={2015},
  pages={5206-5210},
  doi={10.1109/ICASSP.2015.7178964}}

@inproceedings{remy24_colm,
  author       = {Remy, François and Delobelle, P. and Avetisyan, H. and Khabibullina, Alfiya and de Lhoneux, M. and Demeester, Thomas},
  booktitle    = {Proc. Conf. on Language Modeling (COLM)},
  title        = {Trans-tokenization and cross-lingual vocabulary transfers: language adaptation of {LLM}s for low-resource {NLP}},
  year         = {2024},
}

@inproceedings{ecapa2,
  author       = {Thienpondt, Jenthe and Demuynck, Kris},
  booktitle    = {Proc. IEEE Automatic Speech Recognition and Understanding Workshop (ASRU)},
  title        = {{ECAPA2}: a hybrid neural network architecture and training strategy for robust speaker embeddings},
  year         = {2023},
}

@inproceedings{oostdijk-2000-spoken,
    title = "The Spoken {D}utch Corpus. Overview and First Evaluation",
    author = "Oostdijk, Nelleke",
    booktitle = "Proc. Intl. Conf. on Language Resources and Evaluation (LREC)",
    year = "2000",
}

@misc{openasr,
      title={Open {ASR} Leaderboard: Towards Reproducible and Transparent Multilingual and Long-Form Speech Recognition Evaluation}, 
      author={Vaibhav Srivastav and Steven Zheng and Eric Bezzam and Eustache Le Bihan and Nithin Koluguri and Piotr Żelasko and Somshubra Majumdar and Adel Moumen and Sanchit Gandhi},
      year={2025},
      eprint={2510.06961},
      archivePrefix={arXiv},
      primaryClass={cs.CL},
      note={arXiv eprint 2510.06961},
}

@inproceedings{tseng2025evaluationllmsspeechflawed,
      title={Evaluation of {LLM}s in Speech is Often Flawed: Test Set Contamination in Large Language Models for Speech Recognition}, 
      author={Yuan Tseng and Titouan Parcollet and Rogier van Dalen and Shucong Zhang and Sourav Bhattacharya},
      booktitle={Proc. IEEE Automatic Speech Recognition and Understanding Workshop (ASRU)},
      year={2025},
}

@inproceedings{sot,
  title     = {Serialized Output Training for End-to-End Overlapped Speech Recognition},
  author    = {Naoyuki Kanda and Yashesh Gaur and Xiaofei Wang and Zhong Meng and Takuya Yoshioka},
  year      = {2020},
  booktitle = {Proc. Interspeech},
  pages     = {2797--2801},
  doi       = {10.21437/Interspeech.2020-999},
}

\end{document}